\documentclass[12pt,letterpaper,fleqn]{narms}
\usepackage[T1]{fontenc}
\usepackage[latin9]{inputenc}
\usepackage{geometry}

\usepackage{subfigure}
\usepackage{epsfig}
\usepackage{timesmt}

\usepackage{chicaco}
\usepackage{array}
\usepackage{verbatim}
\usepackage{textcomp}
\usepackage{amsmath}

\makeatletter

\providecommand{\tabularnewline}{\\}

\newcommand{\vv}{\vspace*{1.5ex}}

                            \newcommand{\no}{\noindent}
 \newcommand{\bc}{\begin{center}}
 \newcommand{\ec}{\end{center}}
                   \newcommand{\bfr}{\begin{flushright}}
                   \newcommand{\efr}{\end{flushright}}
   \newcommand{\ii}{\item}
     \newcommand{\be}{\begin{enumerate}}
     \newcommand{\ee}{\end{enumerate}}
        \newcommand{\bi}{\begin{itemize}}
        \newcommand{\ei}{\end{itemize}}
            \newcommand{\bd}{\begin{description}}
            \newcommand{\ed}{\end{description}}
                \newcommand{\beq}{\begin{equation}}
                \newcommand{\eeq}{\end{equation}}
                  \newcommand{\bea}{\begin{eqnarray}}
                  \newcommand{\eea}{\end{eqnarray}}

      \newcommand{\bfi}{\begin{figure}}
      \newcommand{\efi}{\end{figure}}
\newcommand{\bay}{\begin{array}{l}}
\newcommand{\eay}{\end{array}}
            \newcommand{\dd}{\mbox{d}}

    \newcommand{\pa}{\partial}

    \newcommand{\al}{\alpha}
    \newcommand{\sig}{\sigma}
    
    \newcommand{\tht}{\theta}  
    \newcommand{\ga}{\gamma}

\newcommand{\nnabla}{\mbox{\boldmath $\nabla$}}
\newcommand{\jj}{\mbox{\boldmath $j$}}

\begin{document}

\thispagestyle{empty}
        \hspace*{1mm}  \vspace*{-0mm}
\noindent {\footnotesize {{\em
\hfill      Posted on ArXiv, submitted to JEM-ASCE} }}
\vskip 1in
\begin{center}
{\large {\bf Extended Microprestress-Solidification Theory (XMPS) for Long-Term Creep and Diffusion Size Effect in Concrete at Variable Environment }
           }\\[20mm]

{\large {\sc Saeed Rahimi-Agham, Zden\v ek P. Ba\v zant, and Gianluca Cusatis }}
\\[1in]

{\sf SEGIM Report No. 18-04/33788r}\\[1.5in]
   
Center for Structural Engineering of Geological and Infrastructure Materials (SEGIM) 
\\ Department of Civil and Environmental Engineering
\\ Northwestern University
\\ Evanston, Illinois 60208, USA
\\[1in]  {\bf April 26, 2018} 
\end{center}

\clearpage   \pagestyle{plain} \setcounter{page}{1}

\title{Extended Microprestress-Solidification Theory (XMPS) for Long-Term Creep and Diffusion Size Effect in Concrete at Variable Environment}
\author{{Saeed Rahimi-Aghdam} \\
{\aff{Graduate Research Assistant, Northwestern University, Evanston, IL.}} \\
\\
{\authornext{Zden\v ek P. Ba\v zant}}\\
{\aff{McCormick Institute Professor and W.P. Murphy Professor of Civil
and Mechanical Engineering }}\\
{\aff{and Materials Science, Northwestern University 2145 Sheridan Road, CEE/A135, Evanston, Illinois}} \\
{\aff{60208; z-bazant@northwestern.edu; corresponding author.}} \\ 
\\
{\authornext{Gianluca Cusatis}}\\
{\aff{Associate Professor of Civil and Environmental Engineering, Evanston, IL.}}}
\date{}

\abstract{The solidification theory has been accepted as a thermodynamically sound way to describe the creep reduction due to deposition of hydrated material in the pores of concrete. The concept of self-equilibrated nanoscale microprestress has been accepted as a viable model for marked multi-decade decline of creep viscosity after the hydration effect became too feeble, and for increase of creep viscosity after any sudden change of pore humidity or temperature. Recently, though, it appeared that the original microprestress-solidification theory (MPS) predicts incorrectly the diffusion size effect on drying creep and the delay of drying creep behind drying shrinkage. Presented here is an extension (XMPS) that overcomes both problems and also improves a few other features of the model response. To this end, different nano- and macro-scale viscosities are distinguished. The aforementioned incorrect predictions are overcome by a dependence of the macro-scale viscosity on the rate of pore humidity change, which is a new feature inspired by previous molecular dynamics (MD) simulations of a molecular layer of water moving between two parallel sliding C-S-H sheets. The aging is based on calculating the hydration degree, and the temperature change effect on pore relative humidity is taken into account. Empirical formula for estimating the parameters of permeability dependence on pore humidity from concrete mix composition are also developed. Extensive validations by pertinent test data from the literature are demonstrated. 
 }
\maketitle

\section{\large Introduction}

Until recently, the microprestress-solidification (MPS) theory \cite{MPS1,MPS2} appeared to give satisfactory predictions of the creep of concrete (long term included) without and with simultanenous drying and temperature changes. 
  In 2014, however, simulations of P. Havl\' asek at Northwestern University (in collaboration with M. Jir\' asek in Prague and with Z.P. Ba\v zant, reported in \nocite{JirHav14}) identified incorrect predictions of the effect of cross section size on the additional creep due to drying, and an excessive delay, behind drying shrinkage, of the additional creep induced by drying. Both deficiencies are here rectified by the extended microprestress-solidification theory (XMPS). Also made are several other amendments intended to improve on previous amendments proposed by Jir\' asek and Havl\' asek (2014) \nocite{JirHav14}. Extensive verification by the existing test data is important, and is an essential objective of this paper.

The solidification theory separates viscoelasticity of the solid constituent, the cement gel, from the chemical aging of the hardened cement paste caused by solidification of gel particles and characterized by the growth of volume fraction of hydration products. This permits considering the viscoelastic constituent as non-aging.  The decrease of compliance is explained by filling of the pores by growing volume fraction of a non-aging constituent, the cement gel, or C-S-H (calcium silicate hydrate) \cite{BazSolid1,BazSolid2}.

The solidification, however, cannot explain the marked decrease of creep viscosities continuing even after the hydration progress becomes feeble. Neither can it explain the drying creep effect (aka the Pickett effect) and the transitional thermal creep. To explain these phenomena, the concept of microprestress was conceived, resulting in the microprestress-solidification (MPS) theory \cite{MPS1}. 
 
The microprestress characterizes self-equilibrated stresses at the nano-scale level. These stresses stretch and break the interatomic bonds resisting the slip of parallel C-S-H sheets and adjacent C-S-H globules, 
which is believed to be the main mechanism of creep in concrete. The microprestress is considered to be the result of disjoining pressures across nanopores filled by adsorbed water layers. Micropresstress cannot be appreciably affected by the applied load and may be imagined as a the effect of a strong but very soft prestressed spring. The micropresstress is initially produced by incompatible volume changes in the microstructure during hydration. It then relaxes but later again builds up when changes of moisture content and temperature create thermodynamic imbalance between the chemical potentials of vapor, liquid and adsorbed water in the nano-pores of cement gel.

The XMPS improvement in the modeling of drying creep, particularly the dependence of macro-scale viscosity on the rate of microprestress, has been inspired
by recent molecular dynamics (MD) simulations of Vandamme et al. \cite{VanBazKet15} and Sinko et al. \cite{Sin-Baz16,SinBazKet18} at Northwestern. They showed that the the viscosity of creep, associated with the rate of relative slip of parallel planar walls (of C-S-H), is greatly diminished by the presence of a water layer between the walls, and that, furthermore, the effective viscosity of slip between the solid surfaces decreases when the water layer moves. An important finding is that the direction of movement of the water layer does not matter, which means that drying and wetting should have a similar effect. While suspected long ago \cite{Baz70,BazChe85}, these facts were not reflected in the original MPS model. They happen to have been one cause of error in drying creep prediction and explain why the creep enhancement due to drying occurs for both drying and wetting.

\section{\large Microprestress-Solidification Theory (MPS)}

Within the service stress range (and with some exceptions for unloading and simultaneous drying), the concrete creep law can be treated as linear in stress and following the principle of superposition in time. Therefore, the creep is fully characterized by the uniaxial compliance function $J(t,t')$, representing the strain in time $t$ caused by a unit sustained uniaxial stress applied at age $t'$. The triaxial generalization need not be discussed here since it is well known how to obtain it under the assumption that the material is isotropic and the Poisson ratio is approximately independent of time (e.g., Ba\v zant(1972) \nocite{Baz72}, (1982)\nocite{Baz82}, Ba\v zant and Jir\' asek(2018) \nocite{CrBook}). 

\begin{figure}
\includegraphics[width=1\textwidth]{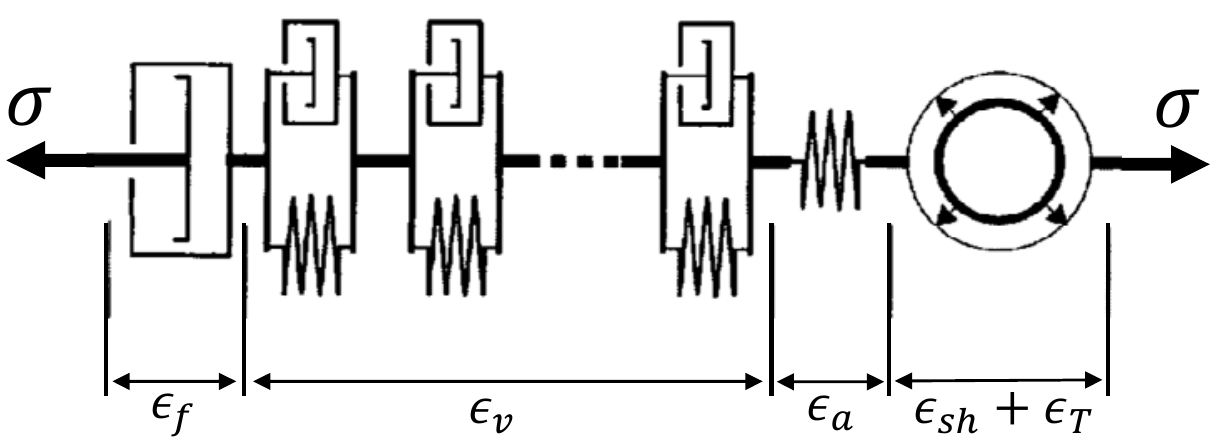}
 \caption{Rheological model. \label{fig:Rheo}}

\end{figure}

In absence of significant plastic strains that may arise at very high confining pressures, the normal strain of concrete can be decomposed as follows (Fig. \ref{fig:Rheo})
 \begin{equation}
  \epsilon = \epsilon_{a} + \epsilon_{v} + \epsilon_{f}
   + \epsilon_{sh} + \epsilon_{T}           \label{eq:totst}
 \end{equation}
where $\epsilon_{a}=$ instantaneous strain; $\epsilon_{v}=$ viscoelastic strain; $\epsilon_{f}=$ flow strain (purely viscous strain); $\epsilon_{sh}=$ shrinkage strain and $\epsilon_{T}=$thermal strain. Not only the flow strain,  but all the concrete creep components are caused by irreversible nanoscale shear slip, occurring mainly between C-S-H sheets at highly stressed sites in the nanostructure (Ba\v zant and Jir\' asek (2018) \nocite{CrBook}). The instantaneous strain, which is the strain appearing immediately after applying uniaxial stress $\sig$, may be written as
 \begin{equation}
  \epsilon_{a} = q_{1}\sigma       \label{eq:Insst}
 \end{equation}
    
Since the retardation spectrum of concrete creep extends smoothly to load durations $\hat t \ll 10^{-4}$ s, it is convenient to define the instantaneous compliance as an asymptotic extrapolation of short-time creep curves for zero load duration (i.e., for $\hat t \to 0$). Such an extrapolation has the advantage that the age effect on the true mean instantaneous compliance happens to be negligible (the evidence for this fact is quite scattered but deviations are not systematic).
This property was demonstrated by Ba\v zant and Osman (1976 \nocite{BazOs76}) and Ba\v zant and Baweja (1995a \nocite{B3p1},1995b \nocite{B3p2}) by considering the measured compliances for load durations $t$ ranging from $0.001$ s to $10$ h. They obtained optimum fits of the compliance values measured for different $t'$  with a smooth formula of the type $J(t,t') = q_{1} + c \hat t^{n}$, where $\hat t = t - t'$ = load duration. Then, by optimizing the fit of the data for various loading ages $t'$, they obtained $q_{1}$ and found that the $q_{1}$ values for various $t'$ were nearly the same. 

Therefore, similar to B3 and B4 models \cite{B3p1,B4Rilem}, the instantaneous compliance (i.e., its asymptotic value for $\hat t \to 0$) is here considered as age independent, which brings about a significant simplification. Introducing an empirical factor $p_1$ depending on the cement type, the instantaneous compliance $q_{1}$ is expressed as:
 \begin{equation}
   q_{1} = \frac{p_{1}}{E_{28}}     \label{eq:q1}
 \end{equation}
where $E_{28}$ is the conventional elastic modulus at age 28 days (which, according to model B4, corresponds to the loading duration of about 0.001 day, or 1.44 min., while in previous model B3 it was 0.01 day).  

The viscoelastic strain $\epsilon_{v}$, which originates in the solid gel of C-S-H, may be described by the same relation type as the B3 and B4 models (Ba\v zant and Baweja 1995, RILEM 2015),
 \begin{equation}
  \epsilon_{v}=\sigma\left\{q_{2}Q(t,t')+q_{3}\ln\left[1
  +(t-t')^{n}\right]\right\}     \label{eq:visst}
 \end{equation}
Function $Q(t,t')$ was derived by asymptotic arguments (Ba\v zant and Prasannan 1988) in a differential form. Its integration leads to a binomial integral that cannot be expressed in closed form. But, in numerical structural analysis in time steps, the integral is not needed, and is even useless if the pore humidity or temperature varies. If $Q(t,t')$ is desired, the integral can be easily evaluated numerically. 

Note that $Q(t,t')$ is age dependent and its value decreases as concrete ages. Previous studies neglected the dependence of $Q(t,t')$ on the growing degree of hydration, $\al(t)$ and simply considered it as a function of loading time $t'$. This degree matters when the temperature or pore relative humidity, $h$, varies. The hydration reaction speeds up as the temperature increases and slows down as $h$ decreases. For normal concrete (without silica fume), the hydration at room temperature virtually stops when $h < 0.75$. In modern concrete the pore humidity can drop as low as 0.65 because of selfdesiccation, and external drying causes nouniform humidity profiles evolving in time. To capture these effects, the actual time $t$ needs to be replaced by the equivalent time $\tht$ that is a function of hydration degree.  To calculate $\tht$, a relation of the same type as in the solidification theory is used and calibrated by data fitting:
 \begin{equation}  
  \tht(\al) = \left[ \frac{0.28}{w/c}\left( \frac{\alpha_{u}}{\alpha} 
   - 1 \right)\right]^{-4/3}    \label{eq:Efftime}
 \end{equation}
where $\al = \al(t)$ is a function of time \cite{Rah17}, $w/c$ = water-cement ratio of concrete mix (by weight); $\alpha_{u}$ = ultimate hydration degree in sealed condition, which is a function of $w/c$; and
 \begin{equation}
  \alpha_u = 0.4 + 1.45(w/c - 0.17)^{0.8}    \label{eq:Ulthyd}
 \end{equation}

Using the equivalent time, one can modify the expression for the evolution of $Q(t,t')$  (Eq. 17 in Ba\v zant and Prasannan (1988) ) by replacing the actual time $t$ with the corresponding equivalent time $\tht$.  
 \beq
  \frac{\dd Q(t,t')}{\dd t} = \left(\frac{\lambda_{0}}{\theta(\al(t))}\right)^{m}\frac{n\zeta^{n-1}}{\lambda_{0}(1+\zeta^{n})}
 \eeq    
where $\lambda_{0}=1$ day, $m=0.5$, $n=0.1$ , and $\zeta=t-t'$. This relation can be integrated to find Q(t,t')
\beq
Q(t,t')=\intop_{t'}^{t}\left(\frac{\lambda_{0}}{\theta(\al(t))}\right)^{m}\frac{n\zeta^{n-1}}{\lambda_{0}(1+\zeta^{n})} \label{eq:Q}
\eeq

The integral $Q(t,t')$ from Eq. \ref{eq:Q} cannot be expressed in a closed form thus should be calculated numerically. An approximate asymptotic matching formula for $Q(t,t')$ was developed in Ba\v zant and Prasannan (1988, Eq. 20). Appendix C gives its generalization to the equivalent time. This explicit formulas very accurate and its use reduces the demand for computer time.

Note that the shrinkage strain $\epsilon_{sh}$ is here understood as  a point-wise eigenstrain, whereas in the B3 and B4 models \cite{B3p1,B4Rilem} it represents the average shrinkage of the whole cross section of a long beam or slab. In addition, contrary to model B4, the autogenous shrinkage is here not separated from the drying shrinkage of the cross section. The shrinkage strain is approximately proportional to the relative humidity change, whether caused by external drying of selfdesiccation. Therefore, in step-by-step analysis, at each integration point of each finite element, the rate of shrinkage strain may be calculated as
 \begin{equation}
\dot{\epsilon}_{sh}=k_{0}\frac{\alpha_{u}-\alpha_{0}}{\alpha-\alpha_{0}}\dot{h}
\end{equation}
where $\dot{h} = \dd h / \dd t$ = rate of humidity change, $k_{0}$ = empirical constant, and we set $\alpha_{0}=0.9\alpha_{set}$. Likewise, the thermal strain rate reads
 \begin{equation}
  \dot{\epsilon}_{T}=k_{T}\dot{T}
 \end{equation}
where $\dot{T} = \dd T /\dd t$ = rate of temperature change and $k_{T}$ = empirical constant. 

To complete the creep law, we finally need the rate of flow strain $\epsilon_{f}$, which is discussed next.

\section{ Evolution of Flow Strain and Microprestress}

\begin{figure}
\begin{centering}
\includegraphics[width=0.5\textwidth]{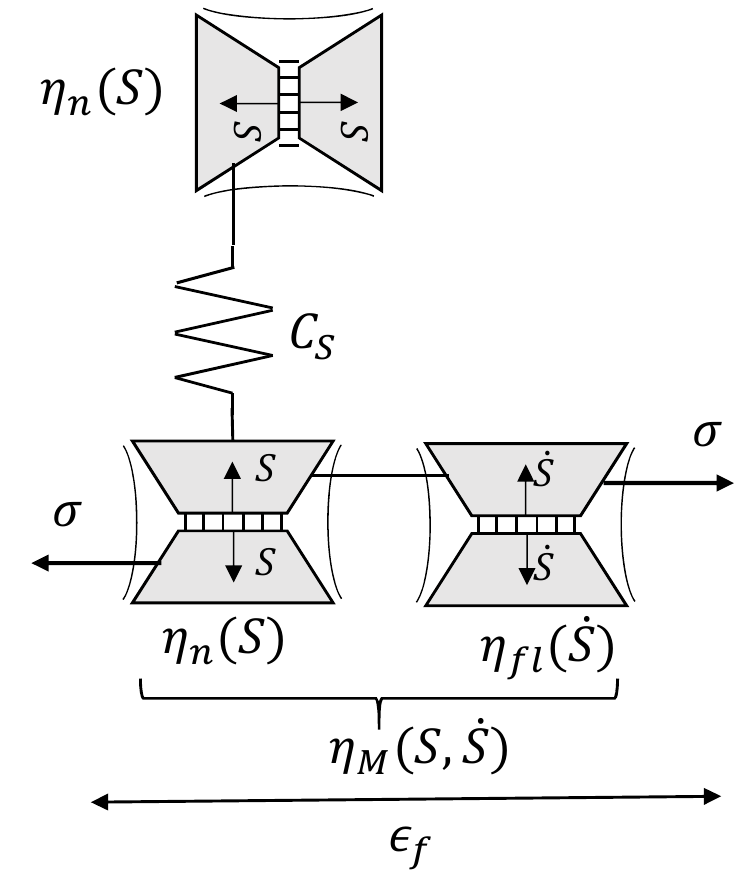}\caption{ Flow strain. \label{fig:MPS}}
\par\end{centering}
\end{figure}

The flow strain is modelled by a viscous flow element coupled in series to the solidifying Kelvin chain, as schematized in Fig. \ref{fig:MPS}. For the flow element portrayed, it is imagined that the bonds across the slip plane cross a nanopore filled by hindered adsorbed water. They are subjected to two stresses: the macroscopic applied stress $\sigma$ causing shear slip, which acts in the figure horizontally, and the tensile microprestress $S$, which acts in the figure vertically. The rate of strain in the flow element is considered to be
 \begin{equation}
  \dot{\epsilon_{f}} = \frac{\sigma}{\eta_{M}}  \label{eq:Flowstrain}
 \end{equation}
where $\eta_{M}$ is the macro-scale viscosity. In the original MPS model, for simplicity, the viscosity is assumed to be the same at the nano-scale and macro-scale, and is in both cases assumed to be a function of microprestress, $S$. This assumption was obviously a simplification since the macro-scale viscosity must depend on several other phenomena as well. In particular, it must depend on the flow of water within the pores (meso-pores or capillary pores, and mainly nano-pores). The water flow accelerates as the rate of pore humidity change increases. 

Recently Sinko et al. (2018) conducted molecular dynamics (MD) simulations of the rate of slip between parallel C-S-H sheets loaded by constant shear stress and by transverse compression. An interstitial layer of water, several molecules thick, was inserted  between the two sheets and forced to move between these two sheets to simulate the flow of water into or out of the nanopore. It was found that the presence of the interstitial water layer accelerates the relative sliding of the C-S-H sheets subject to constant sheer stress and, more importantly, that the rate of relative sliding of the C-S-H sheets accelerates if the interstitial water layer is made to move relative to the mean velocity of the two parallel C-S-H sheets. The explanation is that the movement of the water layer changes the activation energy landscape at the interface, causing the effective viscosity to depend strongly on the flow velocity. The direction of the flow, corresponding to drying or wetting, was found not to be important. 

The conclusion is that the macro-scale viscosity depends not only on the microprestress, but also on the water flow through the pores, the velocity of which is determined mainly by the rate of pore humidity change. It should, of course, be kept in mind that the rate of humidity change is not the only phenomenon that can accelerate the flow between pores. Generally, the rate of any disruption of thermodynamic equilibrium has a similar effect.

Since any phenomenon that causes thermodynamic imbalance increases the microprestress, one can imagine the rate of microprestress as a measure of the water flow rate in the nanopore and the corresponding viscosity to depend only on the microprestress, $S$. In the rheological model of Fig. \ref{fig:MPS}, this viscosity is captured by adding an extra dashpot whose viscosity depends on the absolute value of microprestress rate, $|\dot{S}|$. Considering the absolute value  $|\dot{S}|$ is justified by the MD simulations of Sinko et al. (2018), who found the directions of the flow along a simulated nanopore to be unimportant. As a result of these considerations, the macro-scale viscosity may be introduced in the form, 
 \begin{equation}
  \frac{1}{\eta_{M}} = \frac{1}{\eta_{n}(S)} + \frac{1}{\eta_{flow}}
  = a S + b|\dot{S}|{}    \label{eq:Mvis1}
 \end{equation}
where $\eta_{n}$ is the nanoscale viscosity that is a function of the microprestress only, and $\eta_{flow}$ represents the decrease in viscosity due to the water flow or the rate of any other phenomenon causing thermodynamic imbalance in nanopores. More generally one could consider $\frac{1}{\eta_{M}} = a S^{p_{1}} + b |\dot{S}|^{p_{2}}$ but data fitting indicates no need for such complication.

The viscosity in Eq. \ref{eq:Mvis1} must increase with the age or, more precisely, with the degree of hydration, $\al$, which is a fact confirmed by many experiments \cite{Fer99}. So, it is reasonable to consider both parameters $a$ and $b$ in Eq. \ref{eq:Mvis1} to depend on $\al$ and, for simplicity, a linear dependence can be assumed:
 \begin{equation}
  a = a_{0}\frac{\alpha_{u}}{\alpha},\,\,\,\,\,
  b = b_{0}\frac{\alpha_{u}}{\alpha}       \label{eq:Mvisage}
 \end{equation}
where $a_{0}$ and $b_{0}$ are two empirical constants. Combining
Eq. \ref{eq:Mvis1} and \ref{eq:Mvisage}, we can write
 \begin{equation}
  \frac{1}{\eta_{n}} = \frac{\alpha_{u}}{\alpha}\ a_{0}S,~~~~
  \frac{1}{\eta_{M}} = \frac{\alpha_{u}}{\alpha}\left(a_{0}S
  + b_{0} |\dot{S}| \right)           \label{eq:Visage}
\end{equation}

The next important issue to consider is the evolution of microprestress. The microprestress $S$ is imagined to characterize the average of normal stresses acting across the slip planes with hindered adsorbed water layers between them. The disjoining pressure in these layers, and thus also the microprestress, $S$, is considered to develop first during the initial hardening of cement paste. During the initial rapid hydration, the microprestress builds up mainly as a result of crystal growth pressures and  localized volume changes close to the nanopores. Therefore, during the initial days of fast hydration, $S$ depends mainly on the hydration degree, $\alpha$, and is calculated simply as
 \begin{equation}
  S = S_{0} = c_{0}q_{4}  \,\,\,\,\mbox{for }~~\alpha < \alpha_{0}
 \end{equation}
where $c_{0}$ is an empirical constant; $q_{4}$ is the creep law parameter in models B3 and B4; $S_{0}$ is the initial microprestress, and $\alpha_{0}$ is the hydration degree prior to which the hydration reaction has the dominant control of microprestress. The value $\alpha_{0} = 0.6\alpha_{u}$ can be considered. 

Later, after the volume changes due to hydration have almost ceased, the changes of microprestress are controlled mainly by the changes in the disjoining pressure, which responds with negligible delay to the changes in the capillary tension and surface tension at the same location. According to Ba\v zant et al. (1997a, b), the evolution of microprestress can be assumed to obey a Maxwell-type rheological model with variable viscosity $\eta_{n}(S)$ and stiffness $C_{s}$ 
 \begin{equation}
  \frac{\dot{S}(t)}{C_{S}} + \frac{S(t)}{\eta_{n}(S)}
  =\frac{\dot{s}(t)}{C_{S}}                 \label{eq:Relax}
 \end{equation}
where $\dot{s}(t) /C_{S}$ is the time rate of Maxwell model strain due to any phenomena that may cause thermodynamic imbalance in the microstructure \cite{CrBook}. These phenomena are analyzed next.

\section{ Temperature and Humidity Effects}

The main phenomena affecting $\dot{s}(t) /C_{S}$ are: 1) the temperature change and 2) the humidity change. First the temperature. Its effect is complicated by interference of several physical mechanisms which can be described as follows,
 \begin{enumerate} \setlength{\itemsep}{-1.5mm} 
 \item A temperature increase accelerates the bond breakages and restorations.
 \item The higher the temperature, the faster the chemical process of
cement hydration, and thus the faster the aging of concrete.
 \item The temperature change alters the capillary tension, crystal growth pressure, surface tension, and disjoining pressure, all of which can alter the microprestress and creep rate. 
 \item The temperature changes alters the internal relative humidity, which either increases or decreases the creep rate.
 \item The temperature increase alters the microstructure of C-S-H and of the weaker interfacial transition zone (ITZ), which usually weakens the concrete. 
 \end{enumerate}
The relative pore humidity affects the creep rate and does so mainly by three mechanisms,
 \begin{enumerate}  \setlength{\itemsep}{-1.5mm}
 \item As the relative humidity decreases the viscosity increases and bond breakage decelerates. 
 \item The higher the humidity, the faster is the chemical process of cement hydration and thus the aging of concrete, which reduces the creep rate.
 \item The evolution of humidity changes the capillary tension, crystal growth pressure, surface tension and disjoining pressure, which all change the microprestress. 
 \end{enumerate}

To predict the creep rate correctly, one must consider the effect of all the aforementioned mechanisms.  Except mechanisms 4 and 5 due to temperature change, they have already been considered in the original MPS model \cite{MPS1}. In addition,  These two mechanisms are significant for the case of temperature varying during the experiment. 
Let us begin with considering the effect of temperature on the entire creep law. In models B3 and B4, this effect was considered using a temperature dependent time $t_{T}$ instead of the actual time $t$. Here we use the same idea to calculate the viscoelastic creep (the effect on flow term will be formulated separately);
 \bea
  t_{T} &=& \int_{0}^{t}\beta_{T}(\tau)d\tau     \label{eq:Temptime}
 \\
  \beta_{T}(t) &=&\mbox{exp}\left[\frac{Q_{h}}{R}\left(\frac{1}{T_{0}}
  - \frac{1}{T(t)}\right)\right]                 \label{eq:ArT}
 \\
 \epsilon_{v}(T) &=& \sigma\left(q_{2}Q(t_{T},t'_{T},t'_{eff}) 
  +q_{3}\ln\left[1+(t_{T}-t'_{T})^{n}\right]\right)  \label{eq:VisCrT}
 \eea
where $T$ = absolute temperature; $T_{0}=$ reference temperature, chosen as $T_{0} = 293$ K; $R=$ universal gas constant; $Q_{h}=$ activation energies for the hydration processes (whose values depend on the cement type); and $t'_{T} =$ value of $t_{T}$ at the time of loading. 

The effect of the first temperature related mechanism on the flow strain rate is formulated by decreasing the viscosity, to reflect the acceleration of bond breakage, and by decreasing $C_{s}$, to reflect  the acceleration of bond restorations. Both temperature effects can be described by an Arrhenius type equation. Since the first relative humidity related mechanism also changes the viscosity, both the  temperature and relative humidity influences are considered simultaneously, as follows:
 \bea    \label{eq:VisTh} 
  \eta(T,h) &=& \eta^{T_{0},Sat}/\beta_{\eta}(T,h) 
 \\
  \beta_{\eta}(T,h) 
  &=& \exp \left[\frac{Q_{\eta}}{R}\left(\frac{1}{T_{0}}
  - \frac{1}{T(t)}\right)\right]\left( p_{0} + \frac{1 - p_0}{
  1 + \left(\frac{1-h}{1-h^{*}}\right)^{n_{h}} } \right)
 \\
  C_{s}(T) &=& C_{s}^{T_{0}}/\beta_{C_{s}}(T) \label{eq:CT}
 \\
  \beta_{C_{s}}(T) &=& \exp \left[ \frac{Q_{C}}{R}\left(
   \frac{1}{T_{0}} - \frac{1}{T(t)} \right) \right]  
 \eea
where $Q_{\eta}, Q_{C} =$ activation energies for the viscosity change and $C_{s}$, while the following parameters are considered as  empirical constants: $p_{0}=0.5$, $h^{*}=0.75$ and $n_{h}=2$. For simplicity, one may assume the same activation energy for both the macro- and nano-scale viscosities. In addition, since generally temperature increase accelerates creep, we assume the viscosity decrease to be dominant, and it is found reasonable to set $Q_{C} = Q_{\eta} /2$. 

The effect of the second mechanism, driven by the changes of temperature and relative humidity, is already included through the acceleration of hydration reaction, and no other modification is needed. The third and fourth mechanisms caused by temperature change, and the third mechanism caused by relative humidity, modify the microprestress value through the changes of capillary tension, surface tension, and disjoining (or crystal growth) pressure. 

\begin{figure}

\begin{centering}
\includegraphics[width=1\textwidth]{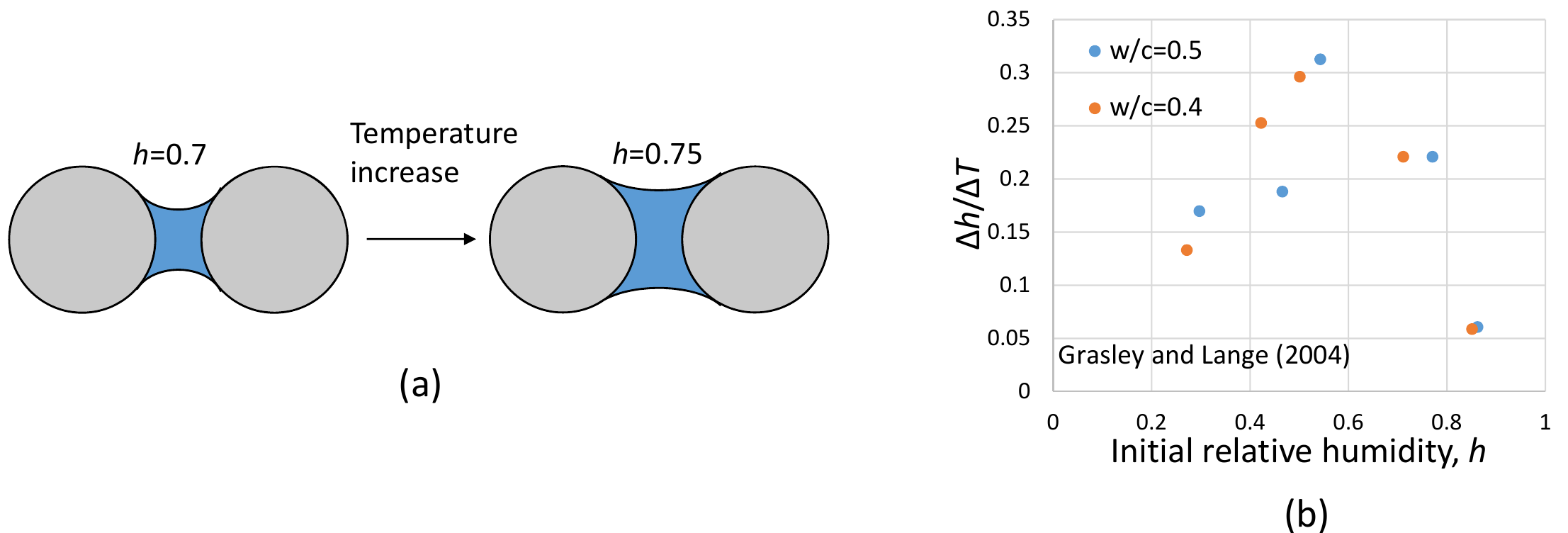}
\par\end{centering}
\caption{ Change of relative humidity due to temperature change. \label{fig:humTemp}}

\end{figure}

All the aforementioned pressures have almost the same relation to the relative humidity, and they are all determined by changes in the chemical potential of pore water ($\mu = (RT/M \rho_l) \ln h$, $\rho_l$ = 1 g/cm$^3$); $\mu$ is, at each location, the same in all the phases of pore water and its change represents the ultimate driving force of hygrothermal deformations. Therefore, and in conformity to the original MPS model (Ba\v zant et al. 1997a,b), all these pressures are combined as one effective pressure, which can be written as
 \begin{equation}   \label{eq:Press}
  p_{eff}=p_{0}+C_{p}\frac{RT}{M}\text{ln}\left(h(t,T)\right)  \end{equation}
where $R=$ universal gas constant, $M$ = molecular weight of water (in moles), $p_{0}$ and $C_{p}$ are two unknown constants. Note that, to capture mechanism 4 due to temperature change, the humidity is here considered to be a function of not only $t$, as in the original MPS model, but also $T$ (this mechanism is considered in Diluzio and Cusatis \citeyear{DilCus13} as well). The rate of effective pressure can then be written as,
 \begin{equation}
  \dot{p}_{eff} = k_{1}\left(\dot{T}\text{ln}\left(h(t,T)\right)
  + \frac{T}{h}\frac{\partial h}{\partial t}
  + \frac{T}{h}\frac{\partial h}{\partial T}\dot{T} 
      \right)                            \label{eq:dotp}
 \end{equation}
where the last term describes the change of relative humidity due to the temperature change (Fig. \ref{fig:humTemp}a). In Eq. (\ref{eq:dotp}), $\pa h /\pa T = \kappa$ = hygrothermic coefficient, introduced in Ba\v zant (1970) \nocite{Baz70} and in Ba\v zant and Najjar, 1972 (Fig. 13) and known to depend strongly on the initial relative humidity at which the temperature change happens; see Fig. \ref{fig:humTemp}b. Here $\kappa$ is calculated based on the experimental results of Grasely and Lange (2007)\nocite{GraLan07}. 
 
The effect of the saturation degree on the effective pressure is not used here directly since the major effects on the microprestress are those of changes in the disjoining pressure and the surface tension, dominant in the smallest pores, which remain filled by water even at very low $h$, such as $h$ > 0.2. This simplification, though, may cause appreciable errors at very low relative humidities.

Next we need to relate these changes to the microprestress change, which is in turn related to the effective pressure change. One can use the simple relation:
 \begin{equation}
  \dot{s}\propto\dot{p}_{eff}
 \end{equation}

The last mechanism that needs to be included is the effect of temperature change on the microstructure of C-S-H and particularly ITZ \cite{Rich91,Cas87,Fah72}. Several studies showed that elevating the temperature at early ages when the high relative humidity is high alters the microstructure of C-S-H, especially in the ITZ, and the average porosity of C-S-H increases. This microstructure change weakens the concrete and can cause sudden increase of microprestres (due to thermodynamic imbalance). One can formulate these effects as follows:
 \bea                                   \label{eq:qT}
  q_{1,2,3}(T) &=& q_{1,2,3}(T_{0})\left[1 + c_{T}(T-T_{0})\right]
 \\
  \dot{s}  &\propto&  f(h)\dot{T}                \label{eq:sT}
 \eea
where Eq. \ref{eq:qT} represents a weakening of microprestress and Eq. \ref{eq:sT} represents its enhancement. Function $f(h)$ introduces the fact that changing C-S-H microstructure gets easier when the relative humidity is high \cite{Fah72,Rich91}. Its simple, empirically calibrated, form is:
 \begin{equation}
  f(h) = c_{h}h^{r}            \label{eq:fh-1}
 \end{equation}
where $r$ is a unknown exponent which is set as $r=3$. 

The reality, though, is a bit more complicated. Fahmi et al. (1972) \nocite{Fah72} showed that the effect of temperature on material stiffness is irreversible and is important if the temperature rises above the range of previously experienced temperatures. Temperature fluctuations within that range do not have much effect.  

Having formulated the mechanisms contributing to the microprestress, we need to combine them. All these contributions are simply assumed to to be independent and additive. Thus we get,
 \begin{equation}
  \dot{s} = c_{p}\dot{p}_{eff} + f(h)\dot{T} 
  = c_{1}\left(\dot{T}\text{ln}\left(h(t,T)\right) 
  + \frac{T}{h}\frac{\partial h}{\partial t} 
  + \frac{T}{h}\frac{\partial h}{\partial T}\frac{\partial T}{
  \partial t} \right) + c_{h}h^{v}\dot{T}          \label{eq:stot}
\end{equation}
where the $c_{p}$ and $c_{1}=k_{1}c_{p}$. Consequently, the equation governing the relaxation of microprestress reads:
 \begin{equation}
  \frac{\dot{S}}{C_{S}(T)} + \frac{S}{\eta_{n}(S,T,h)} 
  = \frac{\dot{s}}{C_{S}(T)}             \label{eq:RelTh}
 \end{equation}
Upon inserting Eqs. \ref{eq:VisTh}, \ref{eq:CT} and \ref{eq:stot} into Eq. \ref{eq:RelTh}, the equation for relaxation of microprestress  becomes:
 \begin{equation}
  \dot{S} + \frac{C_{S}^{T_{0}}}{\eta_{n}^{T_{0},sat}(S)} 
 \frac{\beta_{\eta}(T,h)}{\beta_{C_{s}}(T)}S 
 = c_{1}\left(\dot{T}\text{ln}\left(h(t,T)\right) 
 + \frac{T}{h}\frac{\partial h}{\partial t} 
 + \frac{T}{h}\frac{\partial h}{\partial T}
 \frac{\partial T}{\partial t}\right) + c_{h}h^{r}\dot{T}
 \end{equation}
where $\eta_{n}^{T_{0}, sat}(S) = 1/aS^{p_{1}}$, and $C_{S}^{T_{0}}$ is a constant. As already mentioned, we assume $p_{1} = 1$ and so the microprestress relaxation equation can be simplified as:
 \begin{equation}
  \dot{S} + a_{S}\frac{\beta_{\eta}(T,h)}{\beta_{C_{s}}(T)}S^{2} 
  = c_{1}\left(\dot{T}\text{ln}\left(h(t,T)\right) 
  + \frac{T}{h}\frac{\partial h}{\partial t}
  + \frac{T}{h}\frac{\partial h}{\partial T}
  \frac{\partial T}{\partial t}\right) + c_{h}h^{v}\dot{T}
 \end{equation}
where $a_{S} = C_{S}^{T_{0}}/a$. Finally, using $p_{1}=p_{2}=1$ as   before, we obtain the rate of flow strain: 
 \begin{equation}
  \dot{\epsilon_{f}} = \frac{\sigma}{\eta_{M}(S,T,h)} 
  = \sigma\left(aS + b|\dot{S}|\right)\beta_{\eta}(T,h)
 \end{equation}

\section{ Water Transport (According to Recent Refinement)}

To predict the creep rate under the influence of variable humidity, a realistic model for water transport or drying is obviously important. Over the years, many were proposed. Some were transplants from other porous material, especially soil science, but did not work well since they ignored, or reflected poorly, the particular features of concrete: 1) in concrete, there is a major distributed sink of evaporable water due to continuing hydration; and 2) in normal concrete, neither the vapor phase of water, nor the liquid capillary phase, percolate, and a water molecule moving from one pore to the next must pass through the adsorbed phase in the nanopores.

Because of gradual filling of pores by deposition of hydration products, the hydration sink matters for the pore relative humidity much less than changes in the specific content of evaporable water. Recognizing this, Ba\v zant and Najjar (1972) \nocite{BazNaj72} adopted the pore relative humidity as the primary variable. Their model was incorporated into the Model Code of {\em fib} (F\' ed\' eration internationale de b\' eton, 2013). This model was improved by Cusatis and Diluzio \citeyear{HTC1,HTC2}, based on new experimental evidence. Then it was improved more substantially by Rahimi-Aghdam et al. (2018) \nocite{RahRasBaz18}, in several ways: 1) The humidity dependence of permeability was separated from the diffusivity by using a realistic desorption isotherm; 2) the order-of-magnitude decrease of permeability with decreasing pore humidity was made less steep than it originally was, and was extended below the 50\% humidity; and 3) empirical formulae to estimate the permeability parameters from concrete strength and composition were developed, to make possible realistic estimates without experimental calibration of these parameters. 

Like the Ba\v zant Najjar model, the model of Rahimi-Aghdam et al. (2018) postulates that under constant temperature, the total moisture flux $\jj_w$ is driven by the gradient of pore relative humidity $h$, i.e.,
 \begin{equation}
  \jj_{w} = - c_{p}(h) \nnabla h               \label{eq:1}
 \end{equation}
where $c_{p}$ is the function giving moisture permeability (kg/m.s). 
The condition of mass conservation of water reads:
 \begin{equation}  \label{eq:2}
  \dot{w}_{tot} = - \nnabla \cdot \jj_{w} + \dot{w_{s}}      
 \end{equation}
where $\dot{w}_{s} =$ rate of water mass consumed by the chemical process of hydration, which is calculated from the rate of hydration degree, $\dot \al$, according to the Rahimi-Aghdam et al. (2017) \nocite{Rah17}. Term $\dot{w}_{s}$ represents a distributed sink, leads to self-desiccation, and is particularly important for modern concretes with low $w/c$ or with silica fume, or both. The selfdesiccation and the presence of anticlastic capillary menisci of negative total curvature causes that concrete is never fully saturated, even when pore vapor pressure greatly exceeds the saturation pressure $p_{sat}(T)$ \cite{CrBook}. As it can be seen, the hydration degree plays a great role in several equations. A brief description of model is given in Appendix A (and for details, refer to Rahimi-Aghdam et al., 2017).

For $30\% \ll h \ll 100\%$, the desorption isotherm of pore water may be realistically simplified as follows:
 \begin{equation}
  \dot{h} = k(\alpha,h) \dot w_{t}      \label{eq:3}
 \end{equation}
where $k(\alpha,h)$ (dimension m$^{3}$/kg) is the reciprocal moisture capacity (i.e., the inverse slope of the isotherm), and $\alpha$ is the hydration degree, growing with concrete age. Incorporating \ref{eq:1}, \ref{eq:2}, and \ref{eq:3}, one gets the governing equation of moisture diffusion in concrete: 
 \begin{equation}
  \frac{\partial h}{\partial t} = k(\alpha,h) \nnabla \cdot
  (c_{p}\nnabla h) + \frac{\partial h_{s}}{\partial t}
 \end{equation}
where the last term on the right-hand side represents the self-desiccation sink, which can be calculated using Rahimi-Aghdam et al. (2017) model. The  moisture permeability is calculated by an equation of the same form as in the original Ba\v zant-Najjar model:
 \begin{equation}
  c_{p}(h,\alpha) = c_{1}\left(\beta + \frac{1-\beta}{1 + 
  \left(\frac{1-h}{1-h_{c}}\right)^{r}}\right) \label{eq:perm}
 \end{equation}
where $c_{1}$, $\beta$, $h_{c}$ and $r$ are four empirical parameters. The new model by Rahimi-Aghdam et al. (2018) provides equations to estimate these parameters based on the properties of concrete mix ($a/c$ and $w/c$), which makes experimental calibration of these parameters for a given concrete unnecessary; Appendix B summarizes the equations that are used to calculate permeability. 

\section{ Numerical Simulations and Validation by Test Data}

\begin{center}
\begin{table}
\caption{Common model parameters}

\centering{}%
\begin{tabular}{|c|c|c|c|c|c|c|}
\hline 
Parameter & $a_{0}$ & $c_{h}$ & $Q_{h}/R$ & $C_{S}^{T_{0}}$ & $c_{1}$ & $c_{T}$\tabularnewline
\hline 
value & $0.005q_{4}$ & $0.035$ & $1900$ & $1.6/q_{4}$ & $22.5q_{4}$ & $0.012$\tabularnewline
\hline 
\end{tabular}\label{Tab1}
\end{table}
\par\end{center}

Let us now verify the ability of the current model to reproduce some typical experimental data. In each of the data sets that follow, it is easy to fit one or two curves, and many models can do that. However, it is quite difficult to fit all the curves of a set of diverse experiments on the same concrete with the same material parameters. Achieving such capability has been the objective of this study and is a requirement for a predictive model. Here we demonstrate it for several important data sets from the literature, dealing with various concretes under diverse environmental conditions. 

The available data deal with the basic creep (defined as the creep at no moisture exchange), creep at different temperatures, creep and shrinkage under drying exposure, and transitional thermal creep after a sudden change of temperature. 
Table 1 summarizes the common parameters that were used for all experiments. In addition, the new model by Rahimi-Aghdam et al. (2018) needs only parameters $w/c$ and $a/c$, which were usually reported by the experimenters. The remaining calibration parameters, which are specific to each experiments, are the parameters of long-term creep model, particularly parameters $q_2, q_3, q_4$ and $b_0$ of model B4 (adopted as standard recommendation of RILEM):
 \bea  \label{eq:q2}
  &&q_{2} = c_{2}p_{2}\left(\frac{w/c}{0.38}\right)^{3}
 \\   \label{eq:q3}
  &&q_{3} = c_{3}p_{3}\left(\frac{w/c}{0.38}\right)^{0.4}
   \left(\frac{a/c}{6}\right)^{-1.1}
 \\    \label{eq:q4}
  &&q_{3}=c_{4}p_{4}\left(\frac{w/c}{0.38}\right)^{2.45}
   \left(\frac{a/c}{6}\right)^{-0.9}
 \eea
where $c_2, c_3, c_4$ are three calibration parameters. Based on model B4 paper we set $p_2 = 0.0586$, $p_3 = 0.0393$ and $p_4 = 0.034$. Note that for $c_2 = c_3 = c_4$ the equations become the same as in model B4. This different in optimum values is not surprising because model B4 is was not formulated as a point-wise constitutive law. The last parameter, $b_0$, is simply calculated as:
 \begin{equation}
  b_0=c_{b}q_{4}         \label{eq:b}
 \end{equation}
where $c_b$ is a calibration parameter. Note that here we try to define most parameters as functions of concrete properties and thus minimize the number of unknown parameters to be calibrated by tests. Furthermore, the calibration parameters that we use do not change over a wide range. For instance, $c_2, c_3. c_4$ all vary between 0.6 and 1.5 and for crude estimates can be taken as 1. Thus the XMPS theory model does not require more tedious calibration than the original MPS theory, while its better results are achieved by capturing the underlying phenomena more realistically.   

\subsection{\normalsize Tests of Bryant and Vadhanavikkit (1987)}
  
As mentioned in the Introduction, the impetus for the developing the XMPS was that the original MPS theory predicted an excessive delay of, and a reverse size effect on, the additional creep due to drying, conflicting with the comprehensive tests of Bryant and Vadhanavikkit (1987) \nocite{BryVad87}. Bryant et al. used prismatic and slab specimens of different sizes, both sealed and drying. The concrete had $w/c = 0.47$ and $a/c = 5.1$. During the first two days after casting, the specimens were kept in sealed molds (with no moisture exchange) at temperature $T$ = 293.15 K, and then were exposed for 6 days to an environment of the same temperature and 95\%  relative humidity (RH). Subsequently, groups of specimens were exposed to drying at 60\% RH or sealed, and subjected either to no stress ($\sigma$= 0) or to applied compressive stress $\sigma$ = 7 MPa parallel to the drying surface. The initial strain readings were taken on day 8, before the RH was lowered. All the sealed specimens were prisms 150$\times$150$\times$600 mm. The drying specimens were prisms, of dimensions $D \times D \times$ 600mm, and slabs of the same thicknesses $D$, with sizes $D$ = 100, 150, 200, 300 and 400 mm. 

\begin{figure}

\includegraphics[width=1\textwidth]{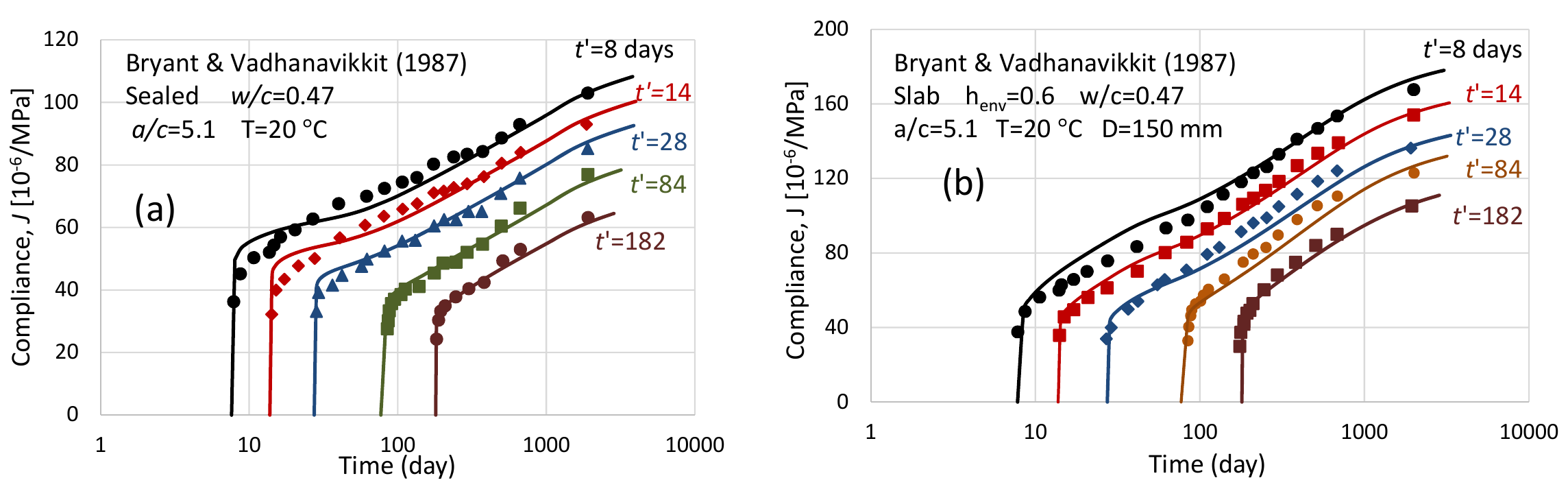}

\caption{ Experimental vs. predicted results for the experiment by Bryant and
Vadhanavikkit (1987).\label{fig:Bry1} }

\end{figure}

First let us examine the prediction of creep at sealed condition, called the basic creep. Fig.\ref{fig:Bry1}a compares the experimental vs. simulated creep values of sealed prisms for different loading times. The calibration parameters are: $c_2 = 1, c_3 = 1, c_4 = 1.1$ and $c_b = 2.5$. As seen, the predictions are satisfactory. 

Next we examine the predictions of creep of exposed specimens, prisms and slabs. For each slab, among the 6 faces the 4 rim faces were sealed and only the two largest faces were exposed (to obtain unidirectional drying across the thickness).  
 
Fig. \ref{fig:Bry1}b compares the experimental and simulated creep values of a drying slab with $D=150$mm, loaded at different ages. As it can be seen, the predicted results are in good agreement. Note that here no extra calibration parameter is used to model the drying process, i.e., the permeability parameters are estimated from Eq. \ref{eq:perm}. 

\begin{figure}

\includegraphics[width=1\textwidth]{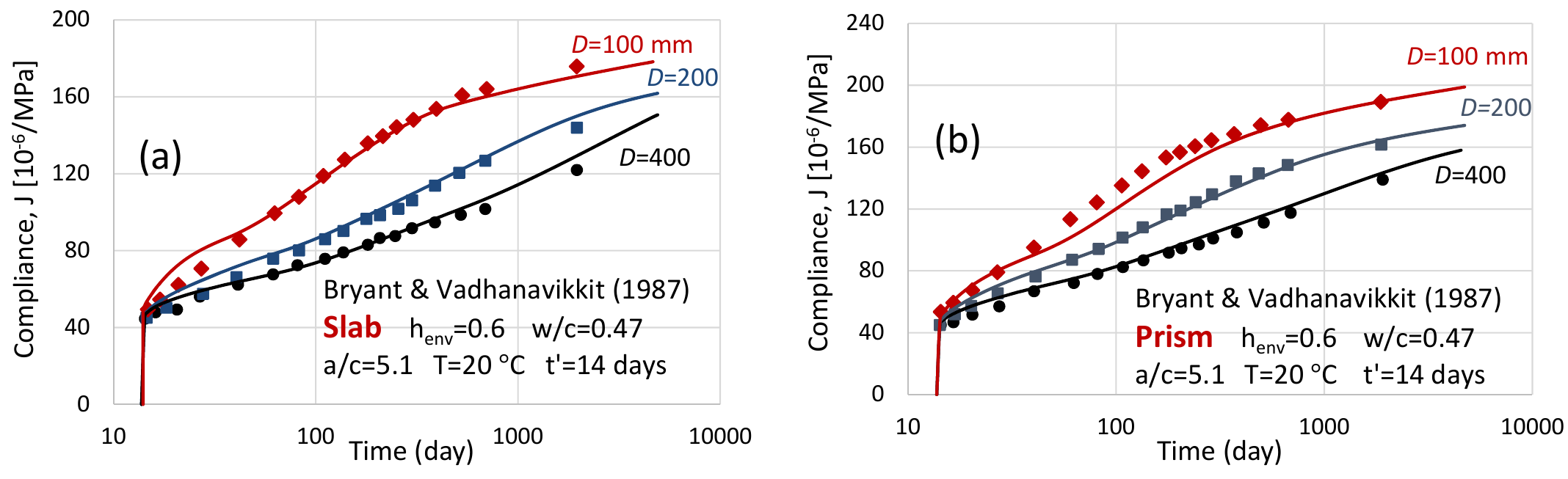}

\caption{ Experimental vs. predicted results for slab and prisms with different
sizes (experiment by Bryant and Vadhanavikkit (1987)).\label{fig:Bry2} }

\end{figure}

After successful prediction of drying creep for the reference size, consider slabs of different sizes (thicknesses). Fig. \ref{fig:Bry2}a demonstrates correct predictions of the diffusion size effect in drying creep, which means that the drying creep in smaller specimens  is faster and the final value of drying creep is bigger. In this regard, note that Havl\' sek at Northwestern found the original MPS model to predict, incorrectly, the opposite---a lower final drying creep for smaller specimens, which contradicted the test results and motivated the development of XMPS. 

Finally we consider the prisms in which only the bottom and top
are sealed and the four long faces were exposed to $h=0.6$. According to the simplification suggested in model B3, the prism can be approximated by an effective cylinder whose diameter is equal 1.09$\times$the side of prism. This approximation, though, did not yield very good results. Therefore, the specimen was simulated in two dimensions as a prism. Exploiting symmetries, only 1/4 of the prism sufficed for analysis. The simulation results using the real prism achieved closer fits (Fig. \ref{fig:Bry2}b). Then, different sizes of effective cylinders were tried and, interestingly, the best approximation occurred when the cylinder diameter was almost equal to the prism side. 

\begin{figure}

\includegraphics[width=1\textwidth]{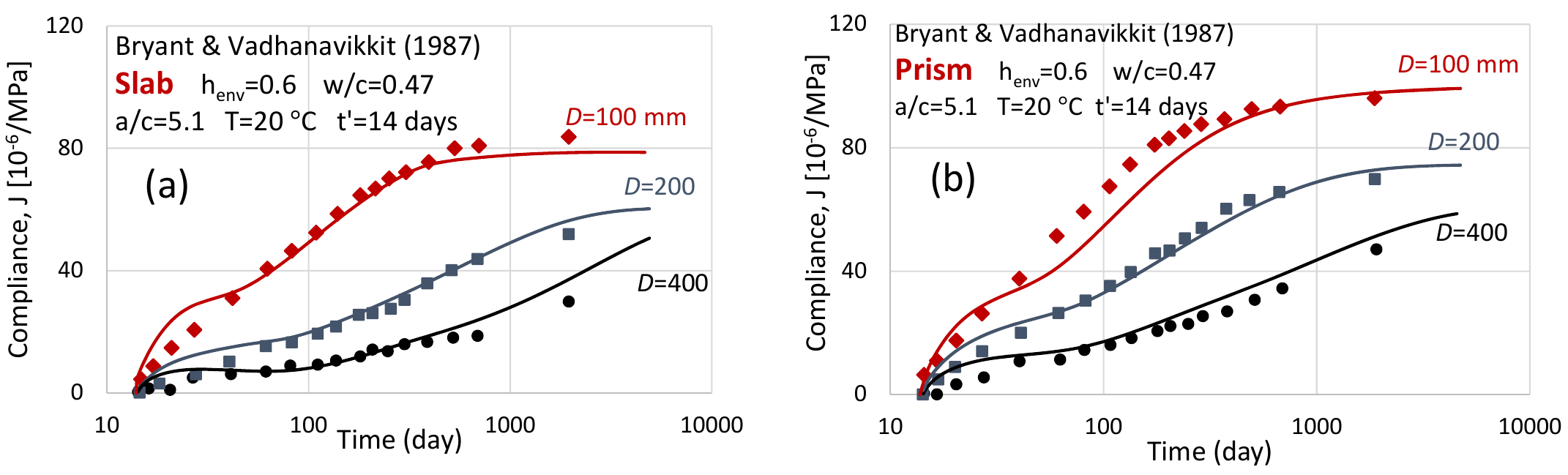}

\caption{ Experimental vs. predicted results for the size effect in drying part of creep. Experiment by Bryant and
Vadhanavikkit (1987).\label{fig:Sizeeff} }

\end{figure}

To illustrate the diffusion size effect on the drying part of creep more clearly, let us subtract the basic creep part. Fig. \ref{fig:Sizeeff} shows the drying part of creep for specimens with different sizes. As it can be seen, the diffusion size effect is significant and the XMPS theory is able to predict it. 

\begin{figure}
\begin{centering}
\includegraphics[width=1\textwidth]{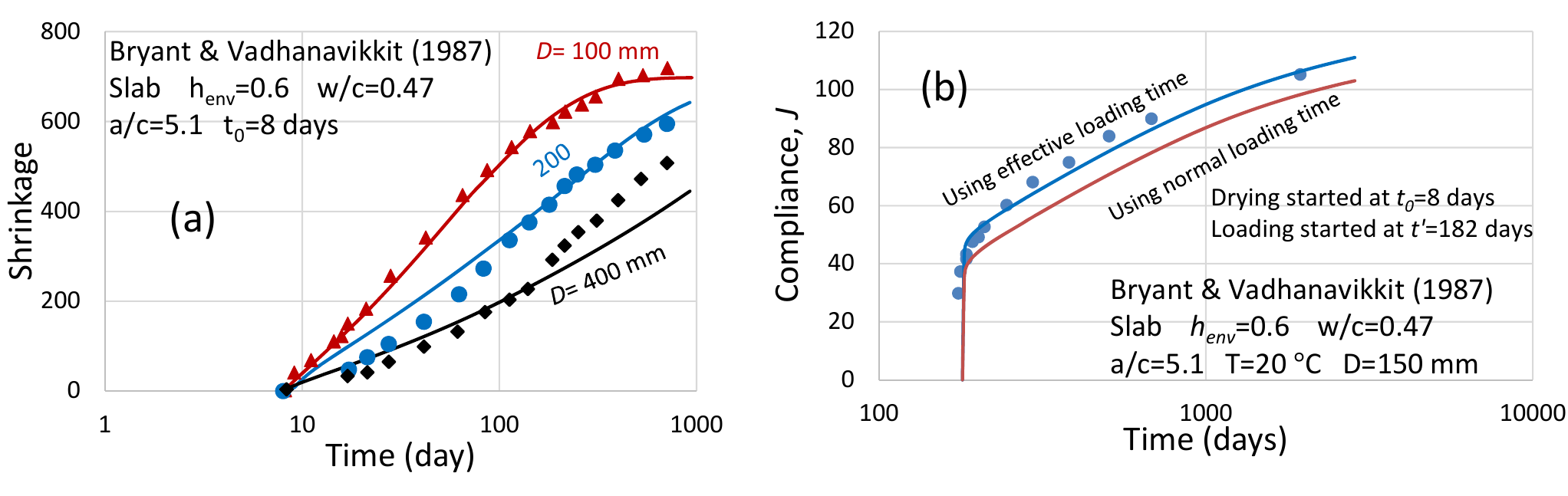}
\par\end{centering}
\caption{ a) Experimental vs. simulated shrinkage, b)Experimental vs. predicted results using normal loading time and effective
loading time. \label{fig:efftime}}

\end{figure}

Until now we have analyzed the diffusion size effect in drying creep  but, of course, the drying shrinkage is size dependent as well. The cause is the differences in the rate of drying of samples with different size.  Fig. \ref{fig:efftime}a illustrates the predicted vs. simulated shrinkage of specimens with different diameters. As seen, for smaller specimens the shrinkage rate is higher and the shrinkage value larger. 

Finally, let us analyze the effect of considering effective time instead of normal time in Eq. \ref{eq:Efftime}. Fig. \ref{fig:efftime}b shows the results obtained on the basis of both the normal loading time and the effective loading time, for an experiment in which the loading and drying times are different. It is seen that better predictions are obtained with the effective loading time. 

Note all the material parameters of the present predictions are the same for all the simulations. This cannot be achieved with other existing models. The same is also true for the fitting of the data for other concretes which follows.

It should be mentioned that often (especially in bending) the cracking damage can make a significant contribution to the Picket effect \cite{BazXi94}, defined as a difference of the total creep of a drying specimen from the sum of the creep of an identical sealed specimen and of the shrinkage of another load-free identical specimen. The reason if that shrinkage of load-free specimen is diminished by cracking while in compressed creep specimens the effect of cracking is minimal or zero. This phenomenon is more important for thin specimens, for early ages and for flexure (used in the tests of \cite{BazXi94}, in which the cracking explained almost a half of the observed Pickett effect). However, in the tests of Bryant et al., the contribution of cracking to the Picket effect was only about 2\%, and thus was neglected in simulations. Nevertheless, the understanding of cracking contribution to the Pickett effect calls for deeper examination in future research.

\subsection{\normalsize Tests of Kommendant et al. (1976) }

Kommendant, Polivka and Pirtz \cite{Kom76} measured creep for different ages $t'$ at loading and at different temperatures. Two almost identical concrete mixes were used. One mix (Berks) was characterized by $w/c = 0.381$ and $a/c = 4.34$. The second mix (York) was almost the same, with $w/c = 0.384$ and $a/c = 4.03$. The test specimens were cylinders 6$\times$16 in. (15.24$\times$40.46 cm), sealed against moisture loss. All the specimens were cured at 23$^\circ$C, and 5 days prior to loading the temperature started increasing at a constant rate of 13.33$^\circ$C/day until the target value 43$^\circ$C or 71$^\circ$C was reached. Furthermore, for each temperature, several tests at different ages of loading were conducted. The calibration parameters were: $c_2 = 0.93, c_3 = 1.85, c_4 = 1.18$ and $c_b = 1.25$. 
    
\begin{figure}
\includegraphics[width=1\textwidth]{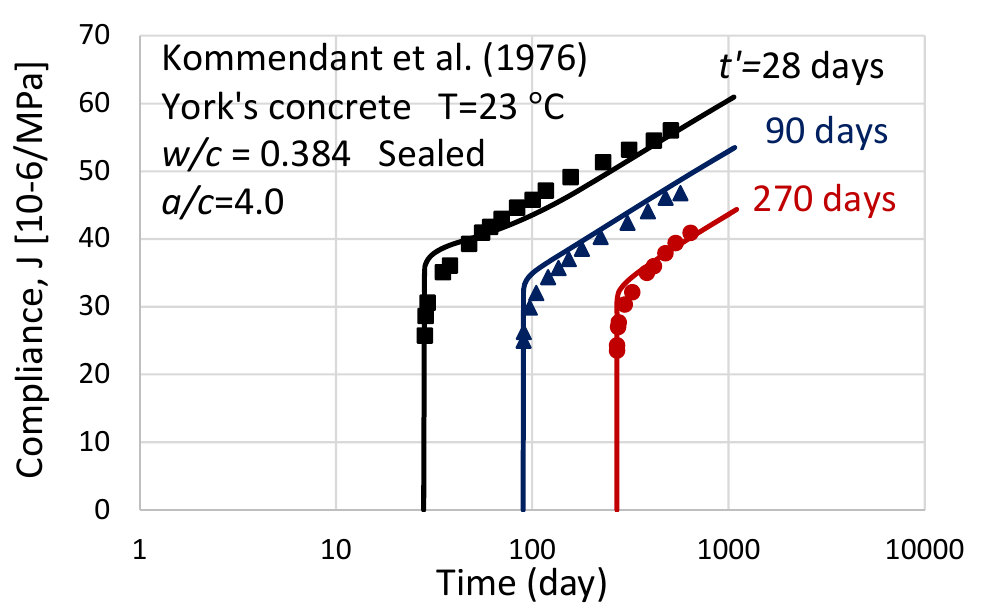}

\caption{ Experimental vs. predicted results of Kommendant et al. (1976) experiment for different loading times at $T=20 ^\circ$C.
\label{fig:Komm}}
\end{figure}

Fig. \ref{fig:Komm} shows the experimental vs. predicted creep values for the York mix at $T=20 ^\circ$C. The results are in good agreement with the experimental ones. Note that although the specimens were sealed, the self-desiccation caused the relative humidity to decrease, which may have affected the microprestress value. 

\begin{figure}
\includegraphics[width=1\textwidth]{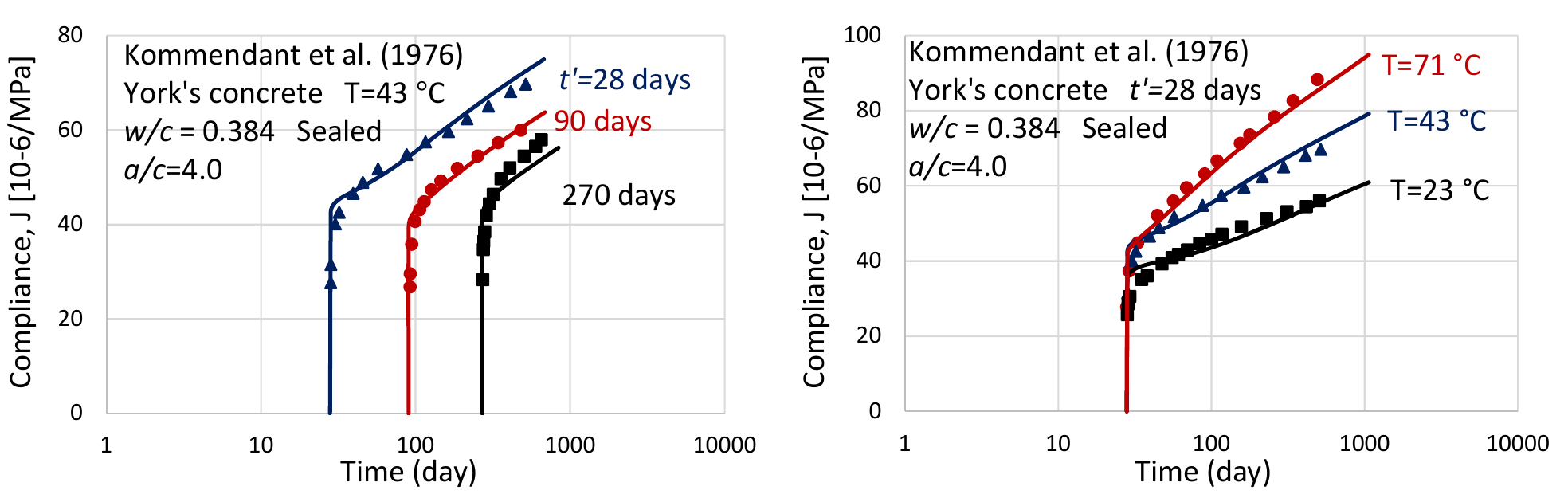}

\caption{ Experimental vs. predicted results of Kommendant et al. (1976) experiment a) different loading times at $T=43 ^\circ$C and b) different temperatures.
\label{fig:Komm2}}
\end{figure}

Next consider the experiments at $T=43 ^\circ$C. Fig. \ref{fig:Komm2}a presents the predicted creep values for different load durations and  for $T=43 ^\circ$C. As can be seen, the predictions agree very well with the experimental data. Note that in the modeling of the experiments at elevated temperatures, increased values of creep parameters $q_{1}, q_{2}, q_{3}$ were considered, due to the temperature induced compliance increase (Eq. \ref{eq:qT}). Finally we examine the ability of the model to predict the creep values for tests with different temperatures but the same loading time. Fig. \ref{fig:Komm2}b documents good accuracy of the model predictions.

\subsection{\normalsize Tests of York et al. (1970) }

\begin{figure}
\includegraphics[width=1\textwidth]{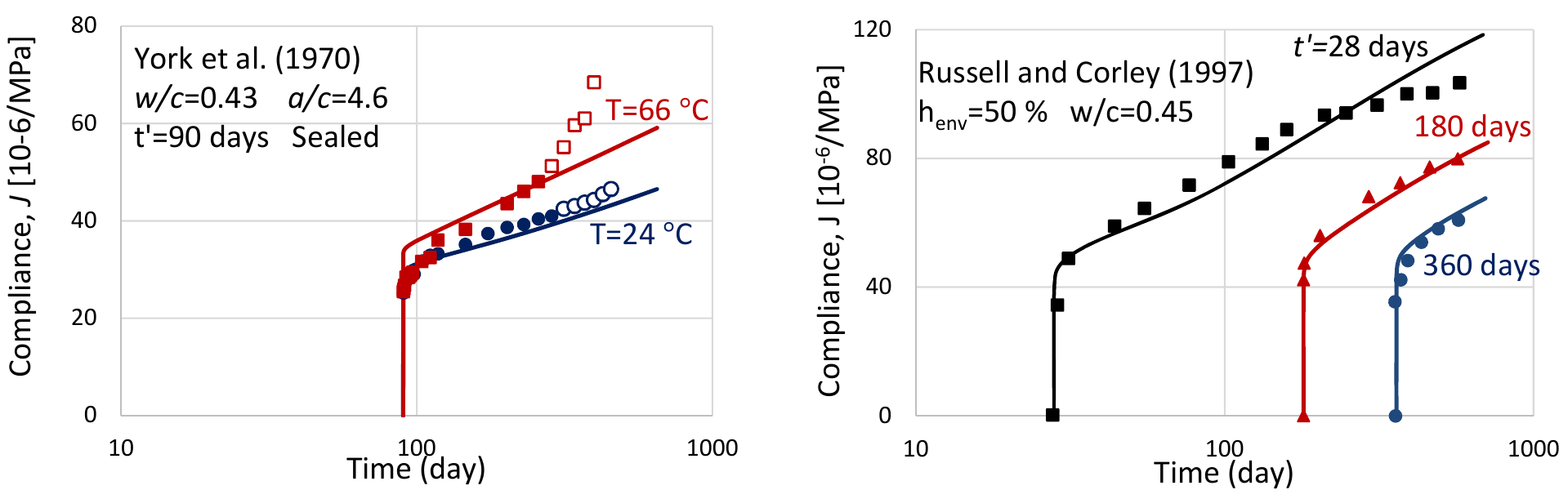}

\caption{ Experimental vs. predicted results of York et al. (1970) and Russell
and Corley (1987). \label{fig:York}}

\end{figure}

York et al. (1970) \nocite{Yor70} conducted several basic creep and drying tests. The creep experiments were carried out on cylindrical specimens of 152 $\times$ 406 cm.  The concrete properties were $w/c=0.43$ and $a/c=4.62$. Unfortunately, after about 300 days the specimens seals failed. Therefore, only the data up to 300 days, shown by solid circles, are fitted, and the subsequent data, shown as empty circles, are ignored. Fig. \ref{fig:York} shows the experimental vs. predicted results. As can be seen, until about $t$=300 days, the predictions are close enough. Note that the same activation energies are here used for all concretes. The calibration parameters are: $c_2=0.8, c_3=1.3, c_4=1.18$ and $c_b=1.25$. 

\subsection{\normalsize Tests of Russell and Corley (1978) }

These drying creep tests \nocite{RusCor78} included three ages of concrete ($t$ = 28, 180 and 360 days). The specimens were cylindrical, with the diameter of 15.2 cm and height of 30.5 cm; $w/c = 0.45$ and aggregate cement ratio $a/c = 3.95$ by weight), exposed to an environment with $h_{env} = 0.5$ and temperature $T = 23$. Drying began after 7 days of curing in a humidity chamber with $h_{env}=1$. Fig. \ref{fig:York}b compares the experimental vs. simulated results for different loading times. The predictions give a close nough agreement with the test data. The calibration parameters were: $c_2 = 1.1, c_3 = 1.1, c_4 = 1.1$ and $c_b = 1.25$. 

\subsection{\normalsize Tests of Di Luzio et al. (2015) }

\begin{figure}

\includegraphics[width=1\textwidth]{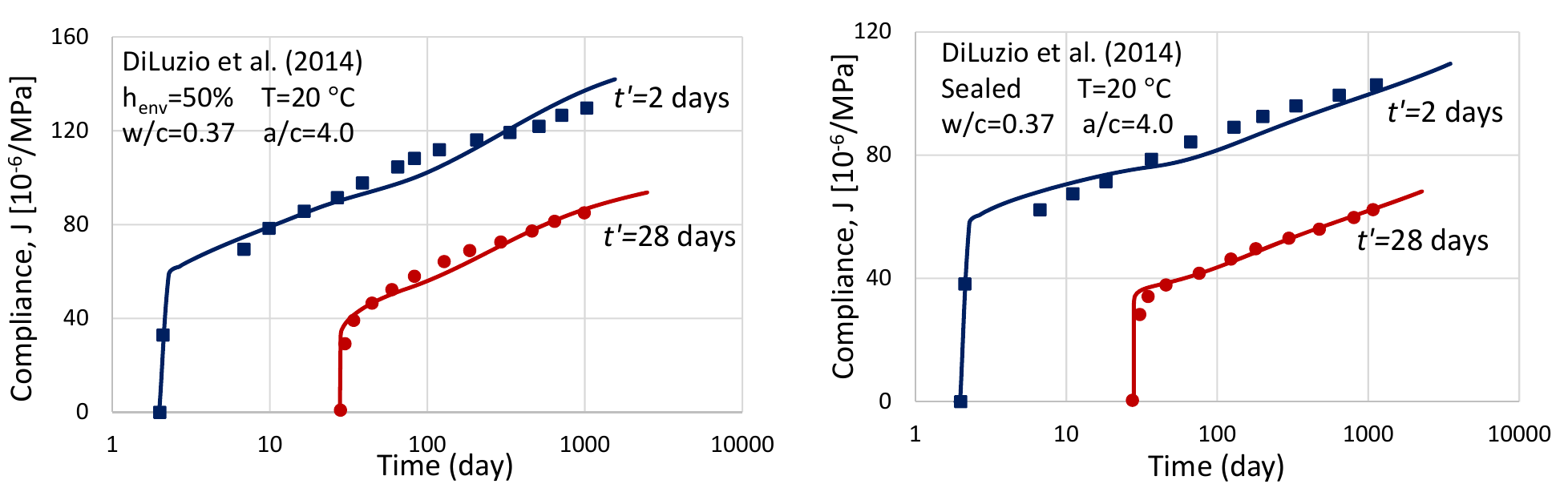}

\caption{Experimental vs. predicted results for the experiments by Di Luzio
et al. (2015).\label{fig:Dil}}
\end{figure}

This experimental study \nocite{Dil15} focussed on basic and drying creep of modern high-performance concrete used in large-span prestressed bridges. In this experimental investigation, cylindrical specimens (of diameter 150 mm and height 360 mm) were used with the environmental relative humidity of 50\% and the temperature of 20$^{\circ}$C. Basic and drying creep tests were conducted starting at the age of 2 and 28 days. The concrete properties were $w/c = 0.37$ and $a/c = 4$. The interesting aspect about these tests is that the loading started at the early age of only 2 days. Fig. \ref{fig:Dil}a shows the experimental vs. predicted results for basic creep. Fig. \ref{fig:Dil}b illustrates the same for drying creep. As can be seen, the results are in good agreement with the test data and indicate that the XMPS theory is able to predict creep correctly even at early ages. The calibration parameters were: $c_2 = 1.35, c_3 = 1.1, c_4 = 1.33$ and $c_b = 1.25$. 
   
\subsection{\normalsize Tests of L'Hermite et al. (1965)}

\begin{figure}

\begin{centering}
\includegraphics[width=0.6\textwidth]{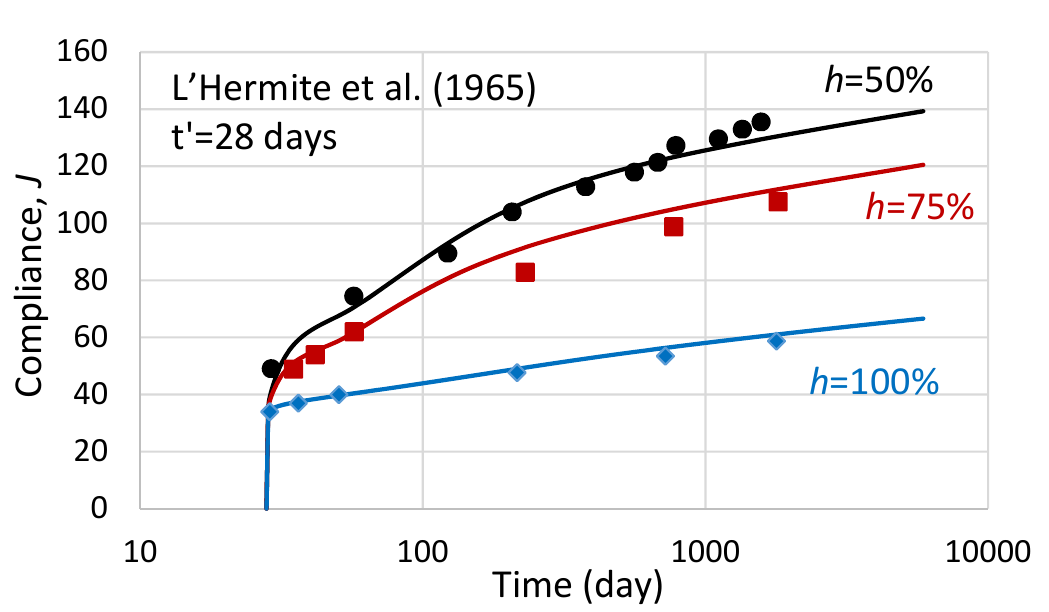}
\par\end{centering}
\caption{ Experimental vs. predicted results for the experiments by L\textquoteright Hermite
et al. (1965). \label{fig:Ler}}

\end{figure}

The comprehensive laboratory study of L\textquoteright Hermite et al.(1965)\nocite{Lher65} included several different types of creep tests. Only the fits of drying creep tests at different humidities are shown here, since good fits of other types of creep tests have already been demonstrated for other data sets. The environmental relative humidities were: $h_{env}$ = 1, 0.75 and = 0.5. The specimens were prisms 28 cm long, with cross section 7 cm $\times$ 7 cm. The concrete mix had $w/c = 0.45$ and $a/c = 3.95$. Fig. \ref{fig:Ler} shows the experimental vs. simulated results for different environmental relative humidity values. Again, the predictions agree well with the experiments. The calibration parameters were: $c_2 = 0.9, c_3 = 0.8, c_4 = 1.0$ and $c_b = 1.25$. 

\subsection{\normalsize Tests of Fahmi et al. (1972)}

\begin{figure}
\includegraphics[width=1\textwidth]{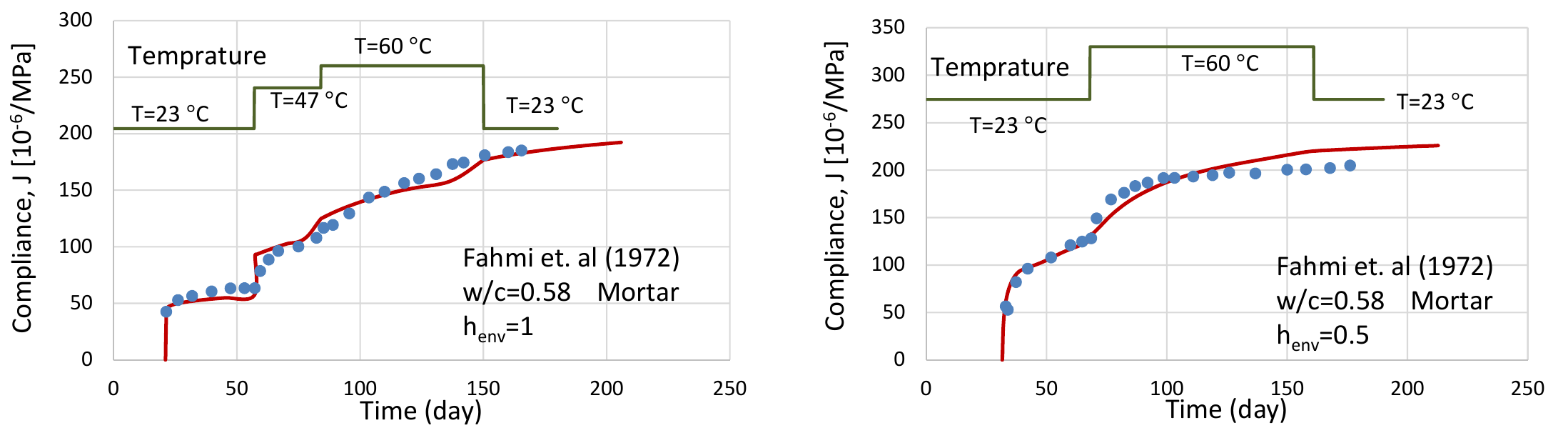}

\caption{ Experimental vs. predicted results for the experiments by Fahmi et
al. (1972). \label{fig:Fahm} }

\end{figure}

In this test series \cite{Fah72}, the temperature was increased during the test. This produced additional creep, called the transitional thermal creep, for both the sealed and drying specimens. The specimens were hollow cylinders with inner diameter 12.7 cm, outer diameter of 15.24 cm, and length of 101.6 cm.  The concrete mix had $w/c = 0.45$ and $a/c = 3.95$. The observed creep curves show upward jumps when the temperature is raised. Fig. \ref{fig:Fahm}a compares the simulations with the data for basic creep and Fig. \ref{fig:Fahm}b does the same for drying creep. 

The predictions are satisfactory. To achieve it, it was important to consider all the mechanisms by which the temperature can change creep rate. Especially, it was important to consider the change of relative humidity in the pores, caused by the temperature rise. Otherwise, the predicted jump for the drying condition would have been much bigger than observed. In addition, it was important to consider the compliance increase due temperature increase (30\% compliance increase of sealed condition was reported). The calibration parameters were: $c_2 = 0.73, c_3 = 1.2, c_4 = 0.8$ and $c_b = 1.25$. It should be mentioned that these experiments were previously by Ba\v zant and Cusatis (2004) \nocite{BazCus04} using a different method.

\section{\large Conclusions}

 \be   \setlength{\itemsep}{-1.5mm}
 \ii
The extended microprestress-solidification theory (XMPS) eliminates two main drawbacks of the original version: 1) the reverse diffusion size effect on the part of creep due to drying, and 2) the excessive delay of the drying part of creep after the drying shrinkage. Both phenomena are now predicted correctly. Predictions of the diffusion size effect on shrinkage are also improved.
 \ii
Different nano- and macro-scale viscosities are now distinguished. The flow term is in XMPS considered to be a function of the macroscale viscosity,  which depends on the pore humidity rate, a feature that transpired from earlier MD simulations showing that the apparent viscosity in sliding of two parallel C-S-H sheet should depend on the velocity of molecular layer of water moving between these two sheets.
 \ii
In the XMPS, the age effect on creep at variable humidity and temperature histories is based not on an empirical effective age (or maturity) but on the effective hydration time calculated from the growth of hydration degree at each point of the structure (or each integration point of a finite element program).  
 \ii
The temperature change effect on pore relative humidity can be realistically described by a humidity dependent hygrothermic coefficient (introduced in \cite{BazNaj71}). 
 \ii
Empirical formulae for estimating the parameters of the humidity dependence of permeability are developed. They mostly obviate the need for calibrating the permeability law by tests.
 \ii
The XMPS is fully compatible with the model B4. Unlike the previous version, all the material parameters can be estimated from model B4, making calibration unnecessary in most applications.
 \ee

\vv {\small \no {\bf Acknowledgment:}\, Partial financial support from the U.S. Department of Transportation, provided through Grant 20778 from the Infrastructure Technology Institute of Northwestern University, from the NSF under grant CMMI-1129449, and from  Nuclear Regulatory Commission (NRC) under award Number
NRC-HQ-60-14-FOA-0001 are gratefully appreciated. Thanks are due to Petr Havl\' asek, visiting researcher at Northwestern co-advised by Milan Jir\' asek from CTU Prague, for discovering the discrepancy of the initial MPS theory vs. Bryant and Vadhanavikkit (1987) size effect data. 
    }
		
		\bibliographystyle{chicaco}
\bibliography{RefrenceMPS}

\section*{\large Appendix A: Algorithm for Evolution of Hydration} 

 \begin{enumerate}
   \item 
For the cement paste or concrete with a known water-cement $w/c$
and aggregate-cement $a/c$ ratios, calculate the initial volume fraction of cement $V_{0}^{c}$and water $V_{0}^{w}$:
 \bea  
  && V_{0}^{c}=\frac{\rho_{a}\rho_{w}}{\rho_{a}\rho_{w}
  +\rho_{c}\rho_{w}a/c+\rho_{c}\rho_{a}w/c}
 \\  
  &&V_{0}^{w}=\frac{\rho_{a}\rho_{c}w/c}{\rho_{a}\rho_{w}
  +\rho_{c}\rho_{w}a/c+\rho_{c}\rho_{a}w/c}
 \eea  
where $\rho_w$, $\rho_c$ and $\rho_a$ are, respectively, the specific mass of water (in 1000 kg/m$^3$), of cement (here considered as 3150 kg/m$^3$), and of aggregates (here 1600 kg/m$^3$), for gravel and sand combined).
   \item 
Calculate the average cement particle size (i.e., particle radius) $a_{0}$, based on the cement type. In this study, the Blaine fineness of cement, $f_{bl}$, equal to 350m$^{2}$/kg was considered to correspond to particle radius 6.5$\mu$m. Also, calculate the number of cement particles, $n_{g}$, per unit volume of cement.
 \bea  
  &&a_{0}=6.5(\mu\mbox{m})\ \frac{350}{f_{bl}}
 \\   
  &&n=\frac{V_{0}^{c}}{\frac{4}{3}\pi a_{0}^{3}}
 \eea  
   \item 
Choose a reasonable hydration degree for setting the time $\alpha_{set}$, and the time $\alpha_{c}$ at which the C-S-H barrier will be completed, i.e., the critical hydration degree at which a complete C-S-H barrier will form around the anhydrous cement grains (about 1 day)). For a normal cement with $a_{0}^{n}=6.5\mu$m,
$w/c=0.3$ at $T=20C$ $^{\circ}$C,  the values $\alpha_{set}^{0}=0.05$ and $\alpha_{c}=0.36$ are good approximations. For specimens with  different $T$, $w/c$ and cement type, reasonable values may be calculated as follows:
 \bea   
  &&\alpha_{c}=\alpha_{c}^{0}f_{1}(a)f_{2}(w/c)f_{3}(T)<0.65
 \\      
  &&f_{1}(a)=\dfrac{a_{0}^{n}}{a}
 \\      
  &&f_{2}(w/c)=1+2.5(w/c-0.3)
 \\       
  &&f_{3}(T)=\mbox{exp}\left[\frac{E_{\al}}{R}\left(
  \frac{1}{273+T_{0}}-\frac{1}{273+T}\right)\right]
 \\         
  &&\frac{\alpha_{set}}{\alpha_{set}^{0}}
  =\frac{\alpha_{c}}{\alpha_{c}^{0}}
 \eea    
    \item 
Calculate the volume fraction of cement. $V_{set}^c$, portlandite,  $V_{set}^{CH}$, and gel (C-S-H plus ettringite), $V_{set}^g$. Using these fractions, calculate the radius of the anhydrous cement remnants, $a_{set}$, and the outer radius of C-S-H barrier, $z_{set}$. To describe the chemical reaction of hydration, use the volume ratios: $\zeta_{gc} = 1.52$ and $\zeta_{CHc} = 0.59$ in the following equations:
 \begin{equation}
  \begin{aligned}
 V_{set}^{c}=(1-\alpha_{set})V_{0}^{c}\,\,\,\,\,\,\,\,V_{set}^{CH}=\zeta_{CHc}\alpha_{set}V_{0}^{c}\,\,\,\, V_{set}^{g}=\zeta_{gc}\alpha_{set}V_{0}^{c}
\end{aligned}
 \end{equation}

 \begin{equation}
a_{set}=\left(\frac{V_{set}^{c}}{\frac{4}{3}\pi n_{g}}\right)^{\frac{1}{3}},~~~~z_{set}=\left(\frac{V_{set}^{c}+V_{set}^{g}}{\frac{4}{3}\pi n_{g}}\right)^{\frac{1}{3}}
 \end{equation}
     \item 
In each time step, use the hydration degree $\alpha$ and humidity $h$ from previous step to calculate the water diffusivity $B_{eff}$:
 \begin{equation}
B_{eff}=B_{0}f_{0}(h_{p})f_{4}(\alpha)
\end{equation}
 \begin{equation}
f_{0}=c_{f}+\frac{1-c_{f}}{1+\left(\frac{1-h_{t}^{p}}{1-h^{*}}\right)^{n_{h}}}
\end{equation}

 \begin{subequations} \label{f4}     \begin{alignat}{1}   & f_4(\al)\  =\ \ga\, e^{-\ga}~~\mbox{for}~~ \alpha \leq \alpha^{*}\\   & f_4(\al)\  =\ (\beta /\al_s)^m e^{(\beta /\al_s)^m}
~~\mbox{for}~~ \alpha > \alpha^{*}   \end{alignat}  
 \end{subequations}
where
\begin{equation}
\ga = \left( \frac{\al}{\al_{max}} \right)^m, \,\, \beta = \alpha-\alpha^*+\alpha^{*}\alpha_{s}/\alpha_{max}
\end{equation}

Use $\al_s=0.3$ , $\al_{max}=\al_c/2$, $\alpha^{*}=0.75\al_c$ and $m=1.8$. Furthermore, $c$, $h^{*}$, $n_h$, $c_f$ are empirical parameters. In this study, we set $n_h = 8$, $h^{*} = 0.88$ and $c_f = 0$.
      \item 
In each step, calculate the radius of the equivalent contact-free C-S-H shells, $\hat{z}$, that give the same free surface area as the actual shell radius $z$ would if the shell surfaces were free, with no contacts; 
 \begin{equation}
  \hat{z}_{t}\
   =\ \frac{z_{t}}{1+\left(\frac{z_{t}-a_{0}}{u}\right)^{5}}
 \end{equation}
where $u = a_{0} / 6.4$. This relation has been modified compared to that in Rahimi-Aghdam et al. (2017), so as to represent a multi-decade continuation hydration better (this change has no effect on model performance up to ten years).
      \item 
in each time step, using pore humidity $h$, the unhydrous cement remnant size $a$, and the C-S-H barrier outer radius $z$ from the previous time step, as well as latest diffusivity value, calculate the water discharge $Q_t^1$:
 \begin{equation}
  Q_{t}^{1} = 4\pi a_{t}z_{t}B_{eff}(\alpha,h_{p})\frac{h_{p}-h_{c}}  
  {z_{t}-a_{t}} \left(\frac{\hat{z}^{2}}{z_{t}^{2}}\right)
 \end{equation}
The last factor on the right-hand side, $\frac{\hat{z}^2}{ z_{t}^2 }$,
serves to consider the reduction of C-S-H shell surfaces due to mutual contacts. 
        \item 
In each step, using the calculated water discharge $Q_t^1$, calculate the increment of cement volume, $\dd V_t^c$, of portlandite, $\dd V_t^{CH}$, and cement gel $\dd V_t^{g}$:
 \begin{subequations} \label{eq:newvolumes}
  \begin{align} 
& V_{t+\dd t}^c=V_t^c+\mbox{d}V_t^c=V_t^c-n_gQ_t^1\zeta_{cw}\mbox{d}t
  \\ 
& V_{t+\dd t}^g=V_t^g+\mbox{d}V_t^g=V_t^g+n_gQ_t^1\zeta_{gw}\mbox{d}t
  \\ 
& V_{t+\dd t}^{CH} = V_t^{CH}+\mbox{d}V_t^{CH} = 
  V_t^{CH}+n_gQ_t^1\zeta_{CHw}\mbox{d}t
 \end{align}
 \end{subequations}
where $V_{t} = V(t)$, etc.; $\zeta_{cw}$, $\zeta_{gw}$ and $\zeta_{CHw}$ are, respectively, the volumes of the cement consumed, of the C-S-H gel produced, and of the portlandite produced per unit volume of discharged water. These volume fractions are calculated as: $\zeta_{cw}=\frac{1}{\zeta_{wc},}$, $\zeta_{gw}=\zeta_{gc}\zeta_{cw}$
and $\zeta_{CHw}=\zeta_{CHc}\zeta_{cw}$. 
       \item 
In each time step, use the calculated $\dd V_t^c$ to calculate the increment of hydration degree $\dd \alpha_t$, and the cement particle radius $\dd a_t$:
 \begin{subequations}\label{eq:new_radius} 
   \begin{align}   
  a_{t+\dd t} = a_t+\mbox{d}a_t 
  = a_t +\dfrac{1}{4\pi a_0^2n_g}\mbox{d}V_t^c   
 \\ 
  \alpha_{t+\dd t} = \alpha_t+\mbox{d}\alpha_t=\alpha_t-\dfrac3{4\pi
   a_0^3n_g}\mbox{d}V_t^c  
  \end{align} 
 \end{subequations}
        \item 
At each time step calculate the increment of gel barrier $\dd z_t$
 \begin{equation}
   \mbox{d}z_{t} = \frac{\mbox{d}V_{t}^{g} +    
    \mbox{d}V_{t}^{c}}{4\pi\hat{z}^{2}}~~~~\mbox{for}~~
     \alpha_{t}>\alpha_{c}
 \end{equation}
        \item 
Finally, calculate the selfdesiccation increment of relative humidity, $\dd h_t^s$, of saturation degree, $\dd S_t^{cap}$, and of  inter-particle porosity, $\dd \phi_t^{cap}$;
  \begin{equation}
 \dd h_{t}^{s} = K_{h}\left(\frac{\mbox{d}V_{t}^{c}(\zeta_{bw}
  + \phi_{np}\zeta_{gc}) - \mbox{d}\phi_{t}^{cap}S_{t}^{cap}}
  {\phi_{t}^{cap}}\right)
\end{equation}

where 

 \begin{equation}
 \begin{aligned}
 \mbox{d}\phi_{t}^{cap} = - (\mbox{d}V_{t}^{g} +  
   \mbox{d}V_{t}^{CH}  + \mbox{d}V_{t}^{c}) + (\phi_{gp}-\phi_{np}) 
    \zeta_{gc}
        \end{aligned}
 \end{equation}
where $\phi_{np}$ is the nano-pore part of gel porosity in which the pores are too small to obbey the Kelvin relation. These pores are assumed to be always saturated. In this study, 2/3 of gel pores are assumed to be nano-pores. 

 \end{enumerate}

\section*{\large Appendix B: Algorithm for determining permeability (diffusivity)}

    Based on Ba\v zant-Najjar (1972) model, which has been embodied in the {\em fib} Model Code 2010 (F\' eration internationale de b\' eton, 2013), the
governing moisture diffusion equation for concrete: 
 \begin{equation}
  \frac{\partial h}{\partial t} = k(\alpha,h)\nabla \cdot \left(c_{p}
  \nabla h\right) + \frac{\partial h_{s}}{\partial t}
 \end{equation}
where  $k(\alpha,h)$ (dimension m$^{3}$/kg) is the reciprocal moisture capacity (i.e, the inverse slope of the isotherm), and $\alpha$ is the hydration degree, $c_p$ is the permeability and the last term on the right-hand side is a distributed sink representing the selfdesiccation. Same as in Ba\v zant-Najjar (1972) model, the dependence of moisture permeability $c_{p}$ on $h$ may again be expressed as follows:
 \begin{equation}
  c_{p}(h,\alpha) =  c_{1} \left(\beta + \frac{1 - \beta}{1 + \left(\frac{1-h}{1-h_{c}} \right)^{r}} \right)  \label{eq:perm2}
\end{equation}
 where in Bazant-Najjar model $c_{1}$, $\beta$, $h_{c}$ and $r$ were four unknown parameters which should be determined based on experiments of relative humidity evolution. However, usually, creep and shrinkage experiments don't report relative humidity values and these values and these parameters should be guessed which can cause significant error. To solve this issue recently Rahimi-Aghdam et al. (2018) has proposed some empirical relations to determine these parameters based on concrete admixtures. They set $r=2$ and proposed following relations to calculate other three parameter.

\begin{equation}
  c_{1} = 60 [1+12(w/c-0.17)^2]\, \alpha/\alpha_{u}
 \end{equation}

\begin{equation} \label{eq:h-c}
  \begin{aligned}
 & h_{c} = 0.77 + 0.22\left(w/c - 0.17\right)^{1/2}
  + 0.15 \left(\frac{\alpha_{u}}{\alpha}-1 \right) \,\,\,\mbox{but}~~h_c < 0.99
    \end{aligned}
 \end{equation}

\bea
  &&\beta = c_{f} /c_{1}
 \\
  && c_{f}^{0} = 60[1 + 12(w/c-0.17)^2]\, \alpha/\alpha_{u}
 \\
  && c_{f} = 
    \begin{cases} c_{f}^{0}~~& (h>h_{s})
    \\
    0.1c_{f}^{0} + 0.9 c_{f}^{0} (h / h_s)^4~~& (h < h_{s})
    \end{cases}
 \eea
where $\alpha_{u}$ is the ultimate hydration degree in sealed concrete, which is a function of $w/c$; it is estimated as 
 \begin{equation}
  \alpha_{u} = 0.46 + .95(w/c-0.17)^{0.6}
  ~~~\mbox{but}~~ \alpha_{u} < 1
 \end{equation}

\section*{\large Appendix C}

Based on the definition of effective hydration time, the asymptotic matching approximation from \cite{MPS1} can be generalized as:
 \bea
\begin{aligned}
  & Q(t,t',t'_{eff}) = Q_{f}(t'_{eff})\left(1+\left( 
  \frac{Q_{f}(t'_{eff})} {z(t,t',t'_{eff})}\right)^{r(t'_{eff})} 
  \right)^{-\frac{1}{r(t'_{eff})}}
 \\
  & Q_{f}(t'_{eff})
  =\left[0.086\left(\frac{t'_{eff}}{1\,\mbox{day}}
  \right)^{\frac{2}{9}}+1.21\left(\frac{t'_{eff}}{1\,\mbox{day}}
  \right)^{\frac{4}{9}}\right]^{-1}
 \\
  & z(t,t',t'_{eff}) =  \left(\frac{t'_{eff}}{1\,\mbox{day}}
  \right)^{-0.5}\ln \left[1 + 
  \left(\frac{t-t'}{1\,\mbox{day}}\right)^{0.1}\right]
 \\
  & r(t'_{eff})=1.7\left(\frac{t'_{eff}}{1\,\mbox{day}}\right)^{0.12}+8
    \end{aligned}
 \eea
where $t'_{eff}$ is the effective hydration time at the time of loading, which is calculated from the hydration degree at the time of loading using Eq. \ref{eq:Efftime}.


\end{document}